\documentclass[aps,prb,amsmath,amssymb,twocolumn]{revtex4-1}

\usepackage{graphicx,tabularx}
\usepackage{dcolumn}
\usepackage{bm}

\usepackage{upgreek}
\newcommand{\bise}{Bi$_2$Se$_3$ }
\newcommand{\bisee}{Bi$_2$Se$_3$}
\renewcommand{\figurename}{Fig.~}

\begin{document}

\title[Topological Insulator Thin Films Starting from the Amorphous Phase - \bise as Example]{Topological Insulator Thin Films Starting from the Amorphous \\Phase - \bise as Example}

\author{J. Barzola-Quiquia} 
\email{j.barzola@physik.uni-leipzig.de}

\author{T. Lehmann}

\author{M. Stiller}

\author{D. Spemann}

\author{P. Esquinazi}
\affiliation{Institute for Experimental Physics II,
University of Leipzig, 04103 Leipzig, Germany}

\author{P. H\"{a}ussler}
\affiliation{Division of Thin Films Physics, Institute of Physics, \\Chemnitz University of Technology, 09107 Chemnitz, Germany}

\date{\today}
\begin{abstract}
  We present a new method to obtain topological insulator
  \bise thin films with a centimeter large lateral length. To produce
  amorphous \bise thin films we have used a sequential
  flash-evaporation method at room temperature. Transmission electron
  microscopy has been used to verify that the prepared samples are in
  a pure amorphous state. During annealing the samples transform into
  the rhombohedral \bise crystalline strcuture which was confirmed
  using X-ray diffraction and Raman spectroscopy. Resistance
  measurements of the amorphous films show the expected Mott variable
  range hopping conduction process with a high specific resistance
  compared to the one obtained in the crystalline phase (metallic
  behavior). We have measured the magnetoresistance (MR) and the Hall
  effect (HE) at different temperatures between 2~K and 275~K.  At
  temperatures $T \lesssim 50$~K and fields $B \lesssim 1$~T we
  observe weak anti-localization in the MR; the Hall measurements
  confirm the n-type character of the samples.  All experimental
  results of our films are in quantitative agreement with results from samples
  prepared using more sophisticated methods.
\end{abstract}

\maketitle

\section{Introduction}
\label{introduction}

Topological insulators (TI) are a new class of materials, which were
theoretically predicted to exist in different Bi based materials
\cite{TEO,ZHANG}. Angle-resolved photoemission spectroscopy (ARPES)
has been used to verify the topological surface
states~\cite{HSIE1,CHEN,HSIE2} of such materials. TI have, due to
strong spin-orbit coupling, an insulating bulk energy gap and generate
conducting topologically protected gapless electronic surface states,
which are robust against disorder and magnetic impurities~\cite{LIU}.

These new materials have very interesting properties from the basic
physics point of view and for future applications, for example the
observation of the quantum spin Hall (QSH) effect~\cite{KOENIG} or the
realization of Majorana fermions for the application in topological
quantum computation~\cite{HASAN,QI}. The alloy \bise was
proposed~\cite{ZHANG} as a three-dimensional (3D) TI, and later it was
experimentally confirmed by ARPES~\cite{XIA} that this system possesses a
single Dirac fermion at the surface. 

Diverse investigations of the transport properties in \bise were
done, e.g.~resistance measurements as a function of the temperature
verify the metallic surface states~\cite{KIM-J}, the Hall effect
confirms that \bise is a n-type carrier and the magnetoresistance
shows weak anti-localization (WAL)~\cite{MATSUO-J,KIM-J}.  The WAL
effect is used as evidence that the electrons moving at the surface of
the material are time reversal symmetry protected. In a TI the WAL
effect is a result of the strong spin-orbit coupling, which, in the
presence of defects, puts the backscattering at a minimum when a
magnetic field or a magnetic impurity is not present. Under the
absence of localization, the surface of a TI remains metallic and is
expected to exhibit perfect WAL due to helical surface
states~\cite{HLN80,NKR07}. Therefore 2D WAL was used to verify the
topological nature in several cases~\cite{B12,NDKYHTZ13,CCKHKQC12},
where the temperature dependence of the coherence length $l_\phi$ is
used to prove that the WAL originated from the 2D surface states
($l_\phi\propto T^{0.5}$ for 2D WAL and $l_\phi\propto T^{0.75}$ for
3D WAL)~\cite{AAK82}.

In the last years tremendous efforts were realized in order to produce
this kind of materials as thin films, nanowires,
nanocrystals~\cite{ZHAO} etc., using different techniques such as
molecular beam epitaxy (MBE)~\cite{ZHANGG,WANG}, metal-organic chemical
vapour deposition~\cite{LIN} including mechanical
exfoliation~\cite{CHECK,HSIE} or chemical vapour
transport~\cite{JIAO}. Nevertheless, the preparation of a large area of
\bise films and the possibility to obtain a variety of morphology by
the combination with electron beam lithography techniques is still a
challenge.

The route of the amorphous-crystalline phase is a powerful method
already used to produce high quality thin films of quasicrystalline
AlCuFe~\cite{ROTH}, AlPdMn~\cite{HAB1} and AlPdRe~\cite{HAB2} samples,
whose existence in the phase diagram does not allow variations of more
than 2\% of the components from the correct concentration. To produce
the amorphous films we use a method which allows us to obtain
centimeter-large homogeneous samples. The composition of the target
material shows a maximum aberration of 1\% when compared with the
source material. This technique is called the sequential
flash-evaporation~\cite{HAU} (SFE) and can be easily combined
with electron beam lithography to obtain samples in the desired shapes,
thickness and can also be used to produce hetero structures, e.g.~for
the study of proximity effects.

\section{Experimental Details}
\label{procedure}

The initial material used for the thin film preparation was commercial
\bise millimetre-large pieces \text{(4--7~mm)} from Alfa Aesar with a
purity of 99.999\%. After milling the pieces to 300--500~$\upmu$m in
diameter, the crushed powder was introduced into the flash evaporation
device. The flash evaporation process was done at room temperature
under a nominal pressure of $\approx$ $5\times 10^{-5}$~mbar. The
films for the transport measurements and X-ray analysis were prepared
onto Si-substrates $(5\times 5\times 0.5~\mathrm{mm}^3)$ coated with
an amorphous insulating 150~nm thick Si$_3$N$_4$ film.
\begin{figure}
\includegraphics[width=\columnwidth]{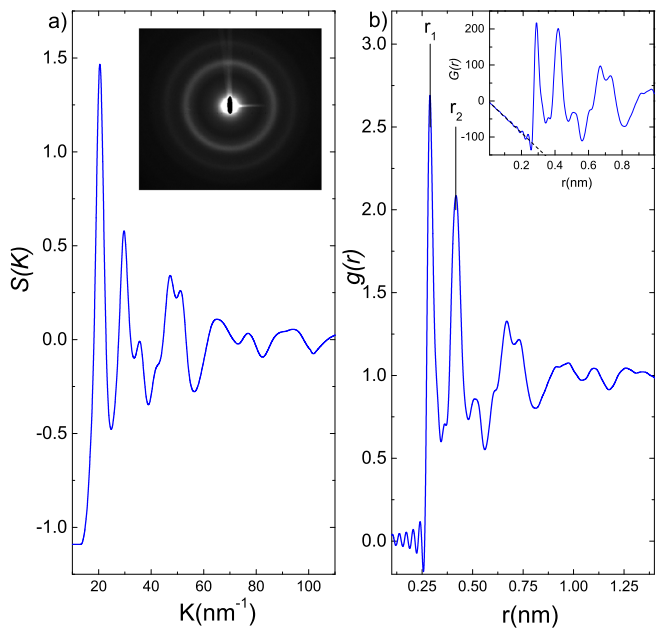}
\caption{\label{fig:skgr} TEM results, a) structure factor $S(k)$ (the
  inset shows the TEM picture), b) pair correlation function
  $g(r)$. The inset shows the reduced atomic distribution function,
  the linear fit at $r < 0.3$~nm is used to obtain the atomic number
  density.}
\end{figure}
\begin{figure}
\includegraphics[width=\columnwidth]{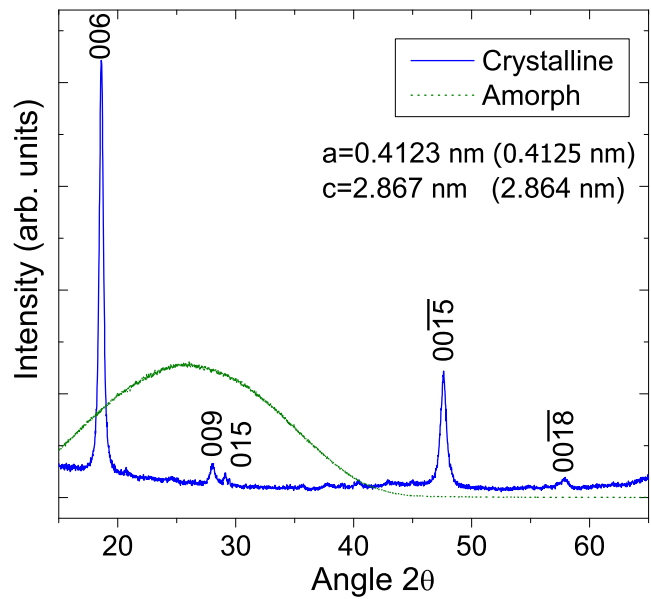}
\caption{\label{fig:xray} X-ray diffraction of the crystalline and amorphous thin
  films. The obtained lattice parameters are in good agreement with the
  literature values (shown in brackets).}
\end{figure}
For transmission electron microscope (TEM) experiments, additional
NaCl crystals substrates were placed into the chamber for deposition,
which, after dilution in bi-distilled water, left the films floating
on the water surface. Thereafter these films were collected onto a
commercial TEM Cu-grid. Atomic structure analysis of the initial
material and the prepared thin films was measured by means of
X-ray diffraction. For TEM measurements we used the transmission electron
microscope Philips CM 20 FEG, to obtain the structure factor
S(\textit{K}) and the pair correlation function g$(r)$, we have
used a standard method described in~\cite{LEITZ}.

The resistance measurements were done using the conventional four
point method; in this case the $5\times 5~\mathrm{mm}^2$ size films
were structured by scratching using a micro-manipulator. The contacts
for low temperature measurements were made out of indium and for the
high temperature measurements we used Pd. The ohmic contacts were
verified by measuring the I-V curves. First low temperature
measurements of the amorphous state were done in a standard commercial
closed cycle refrigerator system with a minimal temperature of
25~K. The electrical resistance was measured using a Keithley 6517A
electrometer due to the high resistance of the amorphous thin films.

Low-temperature measurements between 2~K and 275~K were realized in a
commercial cryostat (Oxford) implemented with a temperature controller
(Linear Research LR-700 Bridge) which allows for a temperature
stabilization of 5~mK over all temperature range.  The
cryostat is also equipped with a $\pm 8$ Tesla superconducting magnet,
which allows to measure the MR and Hall-effect.

Low-noise resistance measurements have been performed using the AC
technique (Linear Research LR-700 Bridge) with ppm resolution. In
order to determine the crystallization temperature we used the DC technique
(Keithley 2182 with 2001 Nanovoltmeter and Keithley 6221 current
source) on the previously structured samples. During the measurements
a magnetic field was applied perpendicular to the sample and current,
and its value was monitored by a Hall sensor installed close to the
sample.

The thickness and stoichiometry of the films were measured with
Rutherford backscattering spectroscopy (RBS)~\cite{CMN78} and particle
induced x-ray emission (PIXE)~\cite{JOH} using a 2.0~MeV proton beam.
There is also an increasing interest in the properties of TI materials
when magnetic impurities are present. Therefore, the amount of
magnetic impurities was quantified using PIXE~\cite{JOH} measurements
as well. This method provides a high sensitivity, for the samples
studied here the typical minimum detection limit is $\approx
2\upmu$g$/$g. The analysis of the RBS and PIXE spectra was performed
using XRUMP~\cite{D78} GeoPIXE II~\cite{RYAN}. The 2.0~MeV protons
used have a penetration depth of $\approx 30 \upmu\mathrm{m}$ in \bise
and therefore PIXE gives the impurity concentration of all the
complete sample if its thickness is less than this value, as it is in
our case.

Raman spectra of the crystalline samples were obtained at room
temperature and ambient pressure with a Dilor XY 800 spectrometer at
514.53~nm wavelength (green) and a 2~$\upmu$m spot diameter. The
incident power was varied between 0.5~mW to 3~mW in order to check for
possible sample damage or laser induced heating effects. No damage or
significant spectral change was observed in this range of incident
power.

\section{Results and Discussion}
\label{results}

After the SFE preparation process at room temperature, the amorphous
samples were obtained. The electron diffraction pattern
of the amorphous sample is shown in the inset of
\figurename~\ref{fig:skgr}. After analysis we obtained the atomic
structure factor $S(K)$ and the pair correlation function $g(r)$, both
results are plotted in \figurename~\ref{fig:skgr}.

The structure results are comparable to known similar amorphous
semiconductor systems~\cite{JB-SC}, but specially to the structure of
the liquid phase of \bise investigated by Usuki~\cite{USUKI}. Our
results show similar features as the liquid phase, e.g.~the position
of the first and second peak in the S$(K)$ and g$(r)$ are almost at
the same values. The first peak at $r_1 \approx 0.29$~nm in g$(r)$
corresponds to the same value in the liquid phase and is close to the
corresponding first nearest neighbor distance in the crystalline phase
of \bisee~\cite{RICHTER}; the peak $r_2 \approx 4.18$~nm was
identified in the liquid phase as consequence of the building of a
short-range order related to the crystalline phase~\cite{USUKI}. After
integration, the first peak in g$(r)$ allows us to calculate the
coordination number $N_c \approx 3.2$. This value is less than that of
the liquid state ($N_{c, \mathrm{Liquid}} \approx 4.12$)~\cite{USUKI}
and is due to the amorphous phase, which builds more defined first
neighbors in comparison to the liquid phase. From the slope of the
reduced atomic distribution $G(r)=4\pi rn_0[g(r)-1]$ at small values
of \textit{r} (dashed line in \figurename\ref{fig:skgr}), we can
calculate the atomic number density $n_0 = 35.05$~nm$^{-3}$, which is
comparable to the value of the liquid phase when calculated using a
simple linear extrapolation of the density of Bi and Se in the liquid
state. From the structure results and the high temperature stability
of the amorphous phase (discussed later), we can conclude, according
to the well established interpretation~\cite{PH-J,JB-J}, that the
system is a stabilized electron-phase, which means the formation of a
pseudogap at the Fermi energy~\cite{PH1-J} is responsible for the high
stability. Consequently, the influence on the electronic transport
yields a very high resistivity and semiconducting-like behavior as a
function of the temperature. After annealing in argon
atmosphere to 425~K, the sample transforms into the crystalline
phase. In \figurename~\ref{fig:xray} we present X-ray results of
the amorphous and crystallized thin films. In case of the crystalline
films we can recognize the peaks corresponding to the rhombohedral
$R\bar 3m$ $(D^5_{3d})$ structure of the \bise phase. The calculated
lattice parameters are in very good agreement with literature
data~\cite{JB-SC} within experimental precision. Using the
Scherrer equation we can determine the crystalline grain size, for
that we used the most intense peak and the calculation gives a value
of $\approx 19 \pm 2.2$~nm. Taking into account that the unit cell is
composed of three layers stacked along the $c$-axis, each known as
quintuple layer (QL), and that a QL has a thickness of $\approx
0.96$~nm, our films contain grains which include 20 QLs. That is
important because in order to have a Dirac cone it is necessary to
have at least 5-6 QL~\cite{ZAHNGGG}.
\begin{figure}
\includegraphics[width=\columnwidth]{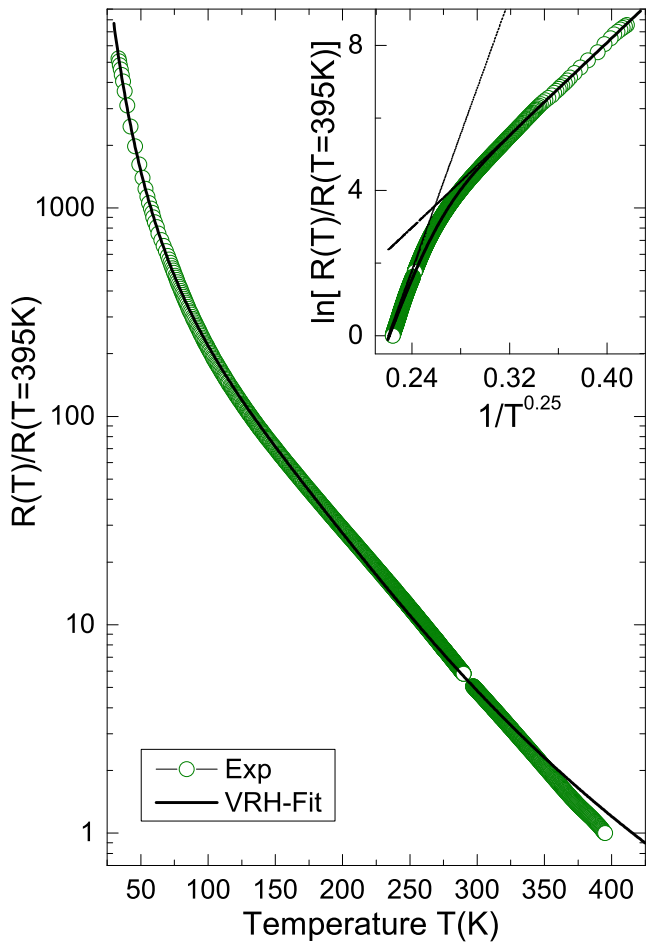}
\caption{\label{fig:tdepamorph} Resistance vs. temperature of the
  amorphous sample after preparation at 320~K. The inset shows the
  fit and the two contributions mentioned in the text.}
\end{figure}
Resistance measurements of the as-prepared amorphous films were also
done, the results are shown in \figurename~\ref{fig:tdepamorph}. In our
device the highest measurable value of the resistance has been reached at
30~K. In the complete temperature range the resistance shows an
thermally mechanism, but with two different regimes, which can be
seen in two different slopes shown in the inset of
\figurename~\ref{fig:tdepamorph}. 
\begin{figure}
\includegraphics[width=\columnwidth]{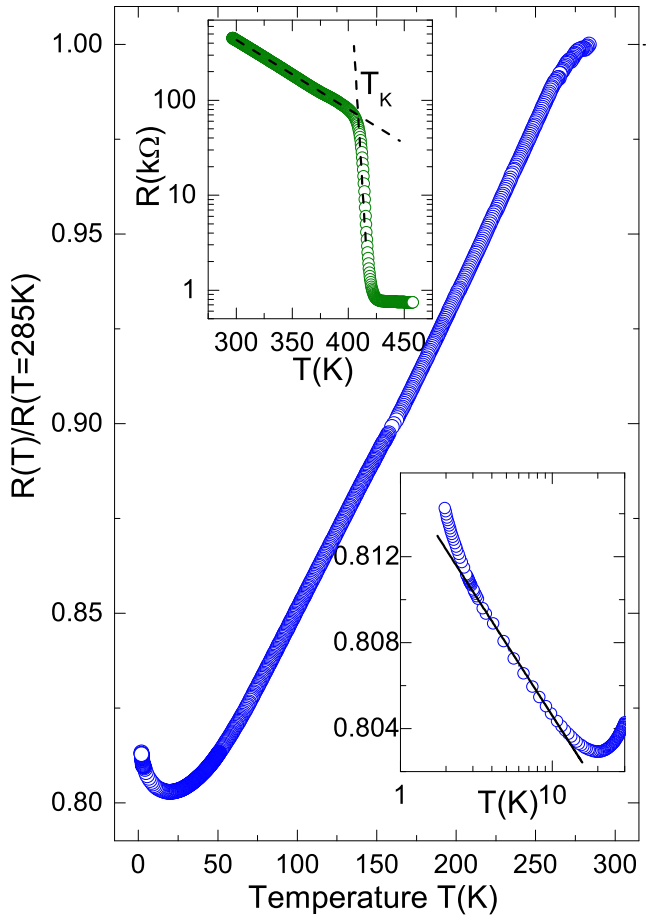}
\caption{\label{fig:r-t} Resistance vs. temperature from 2~K to 275~K
  of the crystalline \bise sample. The upper left inset shows the
  resistance when the sample transforms from amorphous to crystalline
  phase, the intersection of the dash lines gives the crystallization
  temperature $T_{\rm K}$. The inset on the r.h.s.~shows the
  resistance of the crystalline sample.}
\end{figure}
We have found that the temperature dependence of the resistance can be
well described by two different variable-range hopping (VHR) conduction
processes. The general equation for VRH is given by
\begin{equation}
R(T)=R_0\cdot\exp\left[\left(\frac{T_0}{T}\right)^p\right],
\end{equation}
where $R_0$ is a prefactor and $T_0$ is a characteristic temperature
coefficient. The value of the exponent $p$ depends on the nature of
the hopping process. We found that in our amorphous \bise the value of
$p= 0.25$ describes very well the measured transport behavior. This
value of $p$ was derived by Mott~\cite{MOTT}, and that is in the case of a
constant density of states at the Fermi level. According to Mott, $T_0$
is a coefficient which depends on the density of states $N(E_F)$ at the
Fermi level in the form:
\begin{equation}
T_0=\frac{18}{k_B\xi^3N(E_F)},
\end{equation}
where $\xi$ is the localization length and $k_B$ is the Boltzmann
constant.  As we mentioned before, it is evident that the resistance
in our sample shows two different temperature regions, in each one the
resistance can be described by the Mott-VRH. In order to describe the
total temperature dependence we used a simple circuit containing two
resistors in parallel which result is shown in
\figurename~\ref{fig:tdepamorph} as a line.
\begin{figure}
\includegraphics[width=\columnwidth]{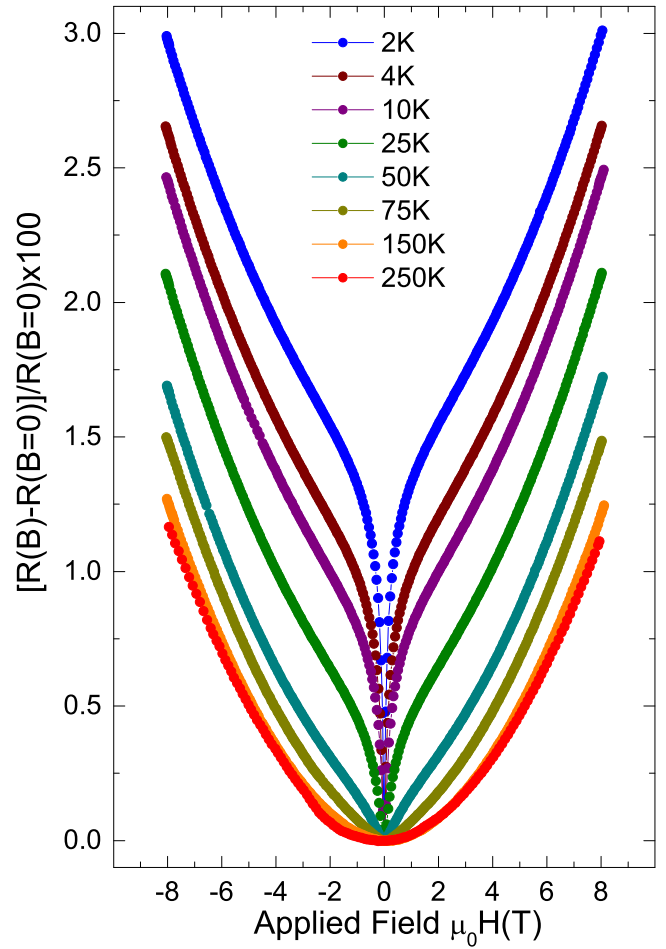}
\caption{\label{fig:MR-T} Magnetoresistance vs. applied magnetic field,
  measured at different constant temperatures.}
\end{figure}
A deviation of the fit from the experimental data at temperatures $T >
320$~K can be observed because the sample was prepared at ambient
temperature and above this value the resistance starts to decrease
irreversibly. The contributions of the two resistances are evident in
the inset. From the fit we obtain the characteristic temperature $T_0$
at every region. For the low temperature region $T_{\rm 0-LT} =
1.02\times10^6$~K and for the high temperature region $T_{\rm 0-HT} =
75.87\times10^6$~K. Using these values we can estimate the
localization length $\xi$ if the value of $N(E_F)$ is known, but this
is not yet available in the literature. Assuming a value of the same
order for similar amorphous systems, $N(E_F)\approx 10^{20(\pm 1)}
\mathrm{eV}^{-1}\mathrm{~cm}^{-3}$, we obtain at low temperatures
$\xi_{\rm LT}\approx 0.5\dots 3$~nm and at high temperatures $\xi_{\rm
  HT} \approx 0.14 \dots0.6$~nm. These values correspond approximately
to the first and second nearest neighbor distances in the pair
correlation function obtained from the TEM investigations. By means of
the VHR analysis it is difficult to conclude on the reason of the change
in the localization length at specific temperature ranges which remains
an open question.
\begin{figure}
\includegraphics[width=.9\columnwidth]{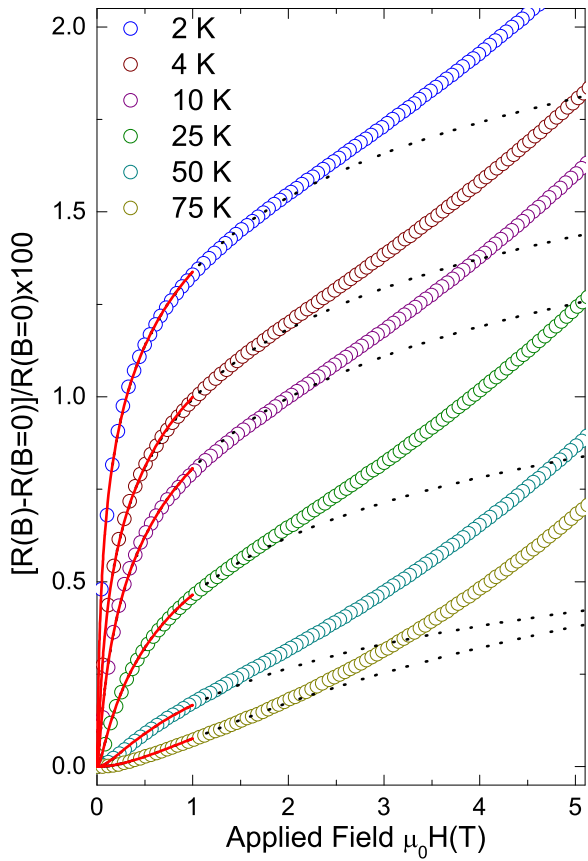}
\caption{\label{fig:MR-T2} Low temperature magnetoresistance, the
  lines show the fits using the WAL theory.}
\end{figure}
\begin{figure}
\includegraphics[width=.9\columnwidth]{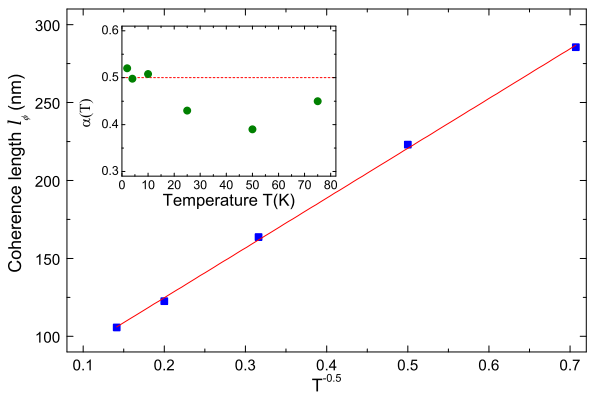}
\caption{\label{fig:koho} Coherence length $l_\phi$(nm)
  vs.~temperature $T^{-1/2} (\mathrm{K}^{-1/2})$ and $\alpha$ (inset)
  obtained from the fits of the magnetoresistance. The red lines in
  the figures are the fits and the expected theoretical values of
  $l_\phi$ and $\alpha$ respectively. }
\end{figure}
The transition from amorphous to crystalline phase was monitored by
the measurement of the resistance during annealing in a home made
device inside an oven under Ar-flow. The results are shown in the
insets of \figurename~\ref{fig:r-t}. The crystallization occurs in a
nearly step-like process and it is evident that the as-prepared film
has a larger resistance in comparison to the crystalline one. The
value of the crystallization temperature $T_K$ was determined by
intersection of the two lines, one before and another one during the
crystallization process. We obtain a $T_K$ of $\approx 410$~K, which
is relatively high in comparison to amorphous metallic
systems~\cite{PH-J,JB1-J} but is in the same range of amorphous-like
materials such as AlCuFe~\cite{ROTH,SCHAUB} and AlPdMn~\cite{HABER,WU}, where
a pseudogap at the Fermi energy was observed by means of angular-resolved
photoemission studies~\cite{WU} and tunneling spectroscopy~\cite{SCHAUB}.

In \figurename~\ref{fig:r-t} we present the results of the temperature
dependence of the resistance of our film in the crystalline phase. At
temperatures between 15~K and 285~K the sample shows the typical
metallic behavior already observed in this system. In the low
temperature range $T < 15$~K we observe an increase in the resistance
with decreasing $T$, which nearly follows a logarithmic temperature
dependence, as observed in some metals~\cite{KHF-J,HE-J} when a small
amount of magnetic impurities is present.  One possible explanation
for this behavior would be an apparent Kondo effect~\cite{KONDO}. Note
that the resistance increases nearly logarithmic below 12~K.  At $T<
3$~K the slope changes which can be a result of the strong influence
of the large magnetic impurity content to the spin-orbit interaction
characteristic for this system.

As consequence of the resistance behavior at low temperature we did
some PIXE measurements. We have analyzed the initial bulk material and
the deposited thin film before and after it was transformed from the
amorphous phase into the crystalline one. The results of the elements
with the highest concentration are summarized in
Table~\ref{table:pixe}. With the exception of Fe, the concentration of
all other magnetic elements was below the detection limit.
\begin{table}[b]
\caption{\label{table:pixe}Results of PIXE measurements.}
\begin{ruledtabular}
\begin{tabular}{p{70pt}p{60pt}p{45pt}p{45pt}}
Sample&Concentration \newline(atom \%)&Fe conc. \newline(in ppm)&Cr conc. \newline(in ppm)
\\ \hline
Initial mat.&Bi$_{\rm 40}$Se$_{\rm 60}$&$34\pm 11$&$<29$\\
Amorphous mat.&Bi$_{\rm 39.4}$Se$_{\rm 60.6}$&$878\pm 53$&$319 \pm 58$\\
Crystallized mat.&Bi$_{\rm 40.2}$Se$_{\rm 59.8}$&$989\pm48$&$339 \pm 59$\\
\end{tabular}
\end{ruledtabular}
\end{table}
\begin{figure}
\includegraphics[width=\columnwidth]{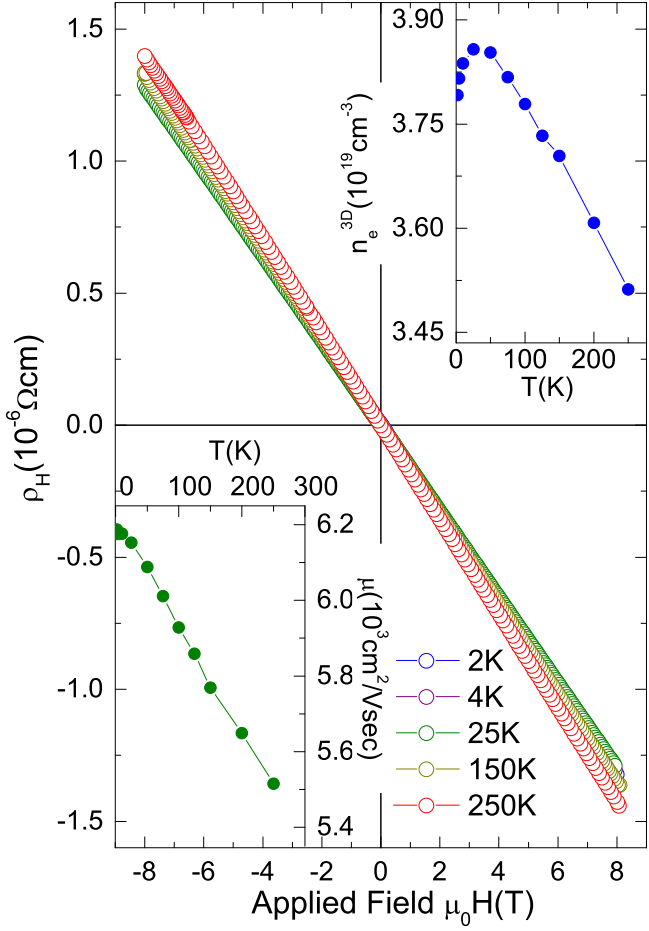}
\caption{\label{fig:hall} Hall resistivity vs. applied field at
  different temperatures. The inset on the l.h.s.~shows the calculated
  mobility, the inset on the r.h.s. the calculated carrier density.}
\end{figure}
According to the PIXE results our preparation method has not produced
considerable change of the initial stoichiometry. This result confirms
the advantages of preparing with the SFE method as before
mentioned. According to PIXE, the Fe (Cr) content already present in
the initial material has increased more than one order of magnitude
after preparation. This contamination can only be due to the milling
process before SFE, because the amount in the amorphous and
crystalline are quite similar, therefore the contamination in the oven
can be ruled out.

In the following we show the results of the transport properties under
the application of a magnetic field. In spite of the magnetic impurity
content and the large polycrystallinity of the film, the results are
quite similar to those reported in the literature for this TI.

The normalized magnetoresistance (MR) is defined as
\begin{equation}
\mathrm{MR} = \frac{R(B)-R(B=0)}{R(B=0)},
\end{equation} 
and are plotted in \figurename\ref{fig:MR-T}. In general the MR is
positive for all measured temperatures and at high fields and high
temperatures ($B > 1.5$~T, $T>100$~K) it is dominated by a parabola
like dependence, due to the Lorenz force contribution to the carriers
scattering under influence of a perpendicular magnetic
field~\cite{HU-J}.
\begin{figure}
\includegraphics[width=\columnwidth]{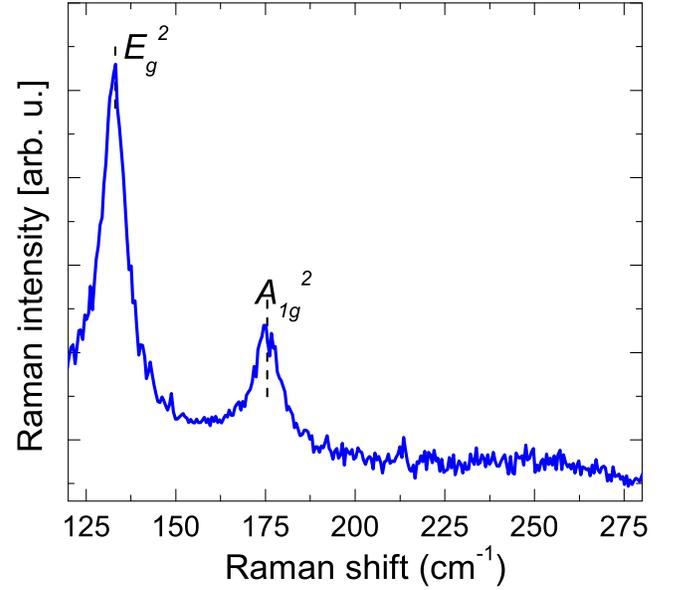}
\caption{\label{fig:raman} Raman spectra of the crystalline \bise
  sample. The dashed lines indicates the characteristic modes.}
\end{figure}
At temperatures $T<75$~K and in the low-field regime ($B<1.5$~T)
the MR increases sharply as the magnetic field increases, which is the
typical behavior of the weak anti-localization (WAL)
effect~\cite{HI-J}. When applying a magnetic field, the time-reversal
symmetry breaks down with increasing field strength and the electrons
begin to backscatter such that the resistance rises. In order to fit
our results with the theory, we define the magnetoconductance (MC) as
$\Delta G(B) = G(B) - G(0)$, with $G(0)=1/R(0)$. The low field
part ($B<1.5$~T) of the MC can be well fitted using the standard WAL
theory developed by Hikami-Larkin-Nagaoka, given as:
\begin{small}
\begin{align}
\label{eq:WAL}
\Delta G_{\rm WAL}(B) &= G(B) - G(0) \nonumber\\ 
&= -\alpha\frac{e^2}{\pi^2\hbar}\left[\psi\left(\frac{1}{2} + \frac{B_\phi}{B} \right) -\ln \left(\frac{B_\phi}{B}\right)\right],
\end{align}
\end{small}
where $\alpha$ is a coefficient whose value was predicted to be 1/2 in
the case of systems with strong spin-orbit interaction, i.e.~in the
pure WAL regime, $\psi$ is the digamma function, $B_\phi$ is the
dephasing magnetic field related to the phase coherence length
$l_\phi$ according to $B_\phi=\hbar/4\mathrm{e}l^2_\phi$. Using the
Eq.~\ref{eq:WAL}, we fit the experimental results, see
\figurename~\ref{fig:MR-T2}, and estimate the different parameters at
each temperature. The prefactor $\alpha$ and the phase coherence
length are shown in \figurename~\ref{fig:koho}. According to our fit
results, at the low temperatures $(T < 20~K)$, $\alpha$ varies around
$0.5\pm 0.2$, which is close to the theoretical expected value. The
high temperature values are between 0.38 and 0.45, this can be a
consequence of the diminishing of the coherence length with rising
temperature. The coherence length varies from 300~nm $(T=2~K)$ to
100~nm $(T=50~K)$, which is in concordance with the value reported for
\bise thin films prepared by molecular beam epitaxy (MBE) with a
thickness of 20 QL~\cite{KIM}.  According to our results, the
temperature dependence of the coherence length is $T^b$, with $b =
−0.51\pm 0.13$. The exponent gives information about the
dimensionality of the system. According to the theory, for a
two-dimensional system, $b = −0.5$ and for one dimension, $b =
−1/3$. Our result is close to the exponent expected for a
two-dimensional system~\cite{CQYLGQZSXYWLL10}. For $\alpha$ and
the $l_\phi$, the observed behavior and values obtained are consistent
with that reported for thin films~\cite{KIM,WANGG,ONOSE} and
microwires of \bise~\cite{MATSUO-J}.

The Hall measurements were done at the same temperatures as the MR and
are presented in \figurename~\ref{fig:hall}. Its sign indicates that
the major carriers are electrons, i.e.~n-type over the investigated
temperature range. From the slope we can calculate the 3D carrier
density, the results are plotted in the inset of
\figurename~\ref{fig:hall}. The value obtained $n_e^{\rm 3D} \approx
3.7 \times 10^{19} \mathrm{cm}^{-3}$ is close to that reported
in~\cite{ANALYTIS} where the sample was grown without dopants but also
to~\cite{CHA} where Fe was used as dopant.

According to theoretical predictions, the presence of magnetic
impurities in a TI should open a small gap in the surface states due
to the breaking of time-reversal symmetry~\cite{LIUU}. To verify these
predictions experimentally, films containing Fe in unknown
concentrations (but with less than 2~\% according to the
authors~\cite{CHA,CCKHKQC12}) were produced. For all doped samples the
results show the typical behavior of the Kondo effect in the
temperature dependence of the resistance. The results of the
magnetoresistance show the typical WAL behavior and only in samples
with a high concentration of Fe, a mix of weak localization (WL) and
WAL effects has been seen~\cite{CCKHKQC12}. Anomalous Hall effect was
observed in \bise thin films when doped with Cr, but only when the
concentration is $\geq$ 1.4~\%~\cite{LIU}. Our Hall effect results do
not show any sign of an anomalous contribution as a consequence of the
Fe and Cr content. Similar results were obtained when \bise was doped
with Fe with a concentration of less than 2~\%~\cite{CHA}. We conclude
that the amount of Fe and Cr present in our sample apparently does not
affect the expected behavior~\cite{LIUU} of the investigated transport
properties of the TI.

Finally, in the crystalline state we have done Raman scattering
spectroscopy of our films. According to the stoichiometry, the
primitive unit cell of \bise contains five atoms, therefore 15 degrees
of freedom are allowed at the center of the Brillouin zone ($q=0$),
from which 3 are acoustic modes and the rest are optical phonons
($2_{\rm A1g}$, $2_{\rm Eg}$, $2_{\rm A1u}$ and $2_{\rm
  Eu}$). According to previous work~\cite{RICHTER}, E$_{\rm g}$ and
A$_{\rm 1g}$ modes can be distinguished. In our films we were able to
detect the high frequency E$^2_{\rm g}$ and A$^2_{\rm 1g}$ modes,
at $\approx 133.14 \mathrm{~cm}^{-1}$ and $\approx 175.5
\mathrm{~cm}^{-1}$, respectively, see \figurename~\ref{fig:raman}.
The measured Raman shift in our samples is in agreement with the
results from literature~\cite{RICHTER, KIM}, indicating that our films
have similar structural quality as others produced with more sophisticated methods.

\section{Conclusion}

TI \bise polycrystalline thin films can be obtained by annealing
starting from the amorphous phase. The preparation was done using a
simple and inexpensive preparation method, which can also be used to
produce similar TI materials, giving the chance to study diverse
properties before, during and after the disorder-order transition. The
material in the amorphous phase has a high resistivity and behaves
like a semiconductor, which transforms in to the metallic crystalline
phase at temperatures higher than room temperature. TEM microscopy,
X-Ray and RAMAN measurements were used to verify that we have produced
the \bise films in the desired crystalline phase. Using PIXE, a method
with high accuracy, we have quantified the presence of magnetic
impurities (such as Fe) in the initial material and in the prepared
thin films. Nevertheless, despite the films being polycrystalline and
that magnetic impurities are present, all measured electronic
transport properties characterizing a TI like \bise were measured and
our results are similar to those reported in literature.  The fact
that our results are quantitatively comparable with data obtained for
samples prepared by more sophisticated methods and of apparently much
better quality, indicates that: either the surface states are
extraordinarily protected and robust against disorder and magnetic
impurities, or the results we are comparing are not the ones that one
should take as characteristics of a good TI.  Future systematic
measurements as a function of disorder and magnetic impurities would
consequently clarify the amount of disorder and magnetic impurities or
both, which causes \bise to become as a non- TI material.


%

\end{document}